\documentclass[a4paper,10pt,twoside]{cpc-hepnp}

\usepackage{multicol}
\usepackage{graphicx}
\usepackage{booktabs}
\usepackage{amssymb,bm,mathrsfs,bbm,amscd}
\usepackage[tbtags]{amsmath}
\usepackage{lastpage}
\usepackage{subfigure}

\begin{document}
%\begin{CJK*}{GBK}{song}

\fancyhead[co]{\footnotesize CHEN Si~ et al: Study on the multipass, multibunch beam breakup for 9-cell TESLA cavities in ERL}

\footnotetext[0]{Received 15 August 2012}

\title{Study on the multi-pass, multi-bunch beam breakup for 9-cell TESLA cavities in ERL\thanks{Supported by the Major State Basic Research Development Program of China under Grant No. 2011CB808303 and No. 2011CB808304.}}

\author{
      CHEN Si(陈思)$^{1,2}$
\quad HUANG Sen-Lin(黄森林)$^{2;1)}$\email{huangsl@pku.edu.cn}
\quad LI Yong-Ming(李永明)$^{2}$
\quad FENG Li-Wen(冯立文)$^{2}$\\
\quad ZHU Feng(朱凤)$^{2}$
\quad QUAN Sheng-Wen(全胜文)$^{2}$
\quad LIU Ke-Xin(刘克新)$^{2;2)}$\email{kxliu@pku.edu.cn}
\quad CHEN Jia-Er(陈佳洱)$^{2}$
}
\maketitle

\address{%
$^1$ School of Physical Science, University of Chinese Academy of Sciences, Beijing 100190, China\\
$^2$ State Key Laboratory of Nuclear Physics and Technology,\\
Institute of Heavy Ion Physics, Peking University, Beijing 100871, China\\
}

\begin{abstract}
Generally, Energy Recovery Linac (ERL) needs special designed high current superconducting RF cavities. In this paper, the threshold current of BBU for compact ERL facilities with 9-cell Tesla type cavities are investigated. The results show that it is feasible to adopt 9-cell Tesla cavity for compact ERL test facilities with just a few cavities and beam current around tens mA.

\end{abstract}



\begin{multicols}{2}

\section{Introduction}

Energy recovery linacs (ERLs) based on superconducting RF technology are suitable for running high current and low emittance electron beam with lower RF power supply than traditional linacs. It's merits indicate a broad prospect of applying ERLs on the next generation light source, high average power FEL, THz radiation and Compton back-scattering facilities \cite{lab1,lab2}.

At Peking University, an ERL test facility, which will operate at 30 MeV and about several milliampere, is under construction. One of the key issues of ERL is the multi-pass, multi-bunch beam breakup (BBU) caused by higher order modes (HOMs) electromagnetic field in RF cavities. It's the main limitation to the available beam current of ERLs. In order to suppress HOMs more efficiently, various types of superconducting cavities have been designed, such as the 5-cell cavity at BNL, 7-cell cavity at Cornell University and 9-cell ERL cavity at KEK/JAEA, etc. Compared with those cavities, the modules of 9-cell Tesla cavities are relatively mature after years of development and some facilities like the International Linear Collider(ILC) and European X-ray FEL have decided to adopt 9-cell Tesla cavity in their main linacs. Although former studies show that 9-cell Tesla cavities may not be applicable for the ERL synchrotron light source which will operate with a current over 100 mA \cite{lab3}, they have the potential to be used in some compact ERLs with just a few cavities and an average current around 10 mA, such as the PKU-ERL test facility. In this paper, we discuss the HOMs and BBU threshold current when 9-cell Tesla cavities are placed in those compact ERLs.

\section{Multi-pass, multi-bunch beam breakup}
Because of the high quality factor of superconducting cavity, HOMs excited by electron bunches may not be sufficiently suppressed. When an electron bunch enters a cavity with excited HOM, it experiences a transverse kick and returns to the cavity with a transverse offset after traveling through the recirculating loop. This offset leads to an energy exchange between HOM and bunch. If the energy gain from bunches is beyond the suppression ability of HOM coupler, HOM energy will grow and larger transverse kicks will be experienced by subsequent bunches, which will in turn lead to further growth of HOM energy. Then, a feedback loop establishes and beam breakup occurs finally.

For the case of single HOM in single cavity, a theoretical equation of BBU threshold current can be expressed as \cite{lab4}

\begin{eqnarray}
\label{eq1}
I_{th} = -\frac{2pc^2}{e\omega(\frac{R}{Q}){Q_e}M_{12}^*\sin({\omega{T_r}})}
\end{eqnarray}

where $\frac{R}{Q}$ is the shunt impedance of HOM; $Q_e$ is the HOM's external quality factor; $\omega$ is the HOM frequency and $M_{12}^*$ is the transport line parameter:

\begin{eqnarray}
\label{eq2}
M_{12}^* = T_{12}\cos^2{\theta}+\frac{1}{2}(T_{14}+T_{23})\sin{2\theta}+T_{34}\sin^2{\theta}
\end{eqnarray}

where $T_{ij}$ is the transport matrix element of the whole transport line; $\theta$ is the polarization angle of HOM. Eq.~(\ref{eq1}) is only available for single HOM in single cavity. For ERLs with more cavities and more HOMs, computer simulations should be adopted. Some codes to calculate the threshold current of BBU have been developed: $TDBBU$, $MATBBU$ and $GBBU$ by Jefferson Lab; $bi$ by Cornell University, etc. It has been proved that all these codes can agree well with the experimental results \cite{lab5}. Here, the code $bi$ \cite{lab6} is used to calculate the threshold current of different cases when 9-cell Tesla cavities are launched in the main linac and the program $elegant$\cite{lab7} is used for particle tracking.

\section{BBU simulation}

\subsection{HOMs in 9-cell Tesla cavity}

According to Eq.~(\ref{eq1}), the most threatening HOMs to BBU should be the dipole modes with larger $(R/Q)Q_e$. Typical simulation results for the 100 mA high-current cavity calculated by Cornell University show that the dipole HOMs should meet the demands of Eq.~(\ref{eq3}) \cite{lab8}

\begin{eqnarray}
\label{eq3}
{(R/Q)Q_e{/f}}<1.4\times10^{5} \Omega/cm^{2}/GHz
\end{eqnarray}

In the 9-cell Tesla cavity, there are several HOMs with $(R/Q)Q_e{/f}>1.4\times10^5$ and they are presented in Table~\ref{tab1}.

\small{
\begin{center}
\tabcaption{ \label{tab1}  4 most threatening HOMs in Tesla cavity}
\footnotesize
\begin{tabular*}{80mm}{c@{\extracolsep{\fill}}cccc}
\toprule $Mode$ & $f$  &   $Q_e$   &   $R/Q$      &   $(R/Q)Q_e/f$        \\
    $No.$ & $GHz$ &  &$\Omega/cm^2$ & $\Omega/cm^2/GHz$ \\
\hline
1 & 1.7074  &   5$\times10^4$   &   11.21   &   3.28$\times10^5$    \\
2 & 1.7343  &   2$\times10^4$   &   15.51   &   1.79$\times10^5$    \\
3 & 1.8738  &   7$\times10^4$   &    8.69   &   3.25$\times10^5$    \\
4 & 2.5751  &   5$\times10^4$   &   23.80   &   4.62$\times10^5$    \\
\bottomrule
\end{tabular*}
\end{center}
}
\normalsize

\subsection{Lattice configuration}

 Lattice configuration should be taken at first. The transport matrix element $T_{12}$ in Eq.~(\ref{eq2}) can be expressed in terms of $\beta$-function and phase advance ${\Delta}\Psi$:

\begin{eqnarray}
\label{eq4}
T_{12}(i\to{f})=\gamma_i{\sqrt{\frac{\beta_i{\beta_f}}{\gamma_i{\gamma_f}}}}\sin{\Delta{\Psi}}
\end{eqnarray}

We assume that all the cavities are fixed in a single cryomodule with no additional focusing between them. The recirculating optics was assumed to be symmetrical which means the recirculating loop (from the end of linac after acceleration to the beginning of linac before deceleration) has equal betatron phase advance in both horizontal and vertical planes and the $\beta$-function to be the same both at the beginning and the end. The transport matrix of RF cavities can be described as \cite{lab9}:

\begin{eqnarray}
\label{eq5}
M_{cav}=
\left(\begin{matrix}\cos{\alpha}-\sqrt{2}\sin{\alpha}&\sqrt{8}\frac{\gamma_i}{\gamma^{'}}\sin{\alpha}\\
-\frac{3}{\sqrt{8}}\frac{\gamma^{'}}{\gamma_f}\sin{\alpha}&\frac{\gamma_i}{\gamma_f}\left[\cos{{\alpha}}+\sqrt{2}\sin{\alpha}\right]\end{matrix}\right)
\end{eqnarray}

where $\alpha=\frac{1}{\sqrt{8}}\ln{\frac{\gamma_f}{\gamma_i}}$, $\gamma_{i(f)}$ is the initial (final) normalized energy of the particle. $\gamma^{'}=qE_{0}\cos(\Delta{\phi})/m_{0}c^2$ is the accelerating gradient of RF cavity.

\subsection{Simulation results}

For an ERL with two 9-cell Tesla cavities, take PKU-ERL test facility for example, 4MeV injected beams will be accelerated to 30 MeV at the first pass. We scanned the betatron phase advance in $0\sim2\pi$ and calculated the BBU current. The results for such a scheme are presented in Fig.~(\ref{fig1})

\begin{center}
\includegraphics[width=8.0cm]{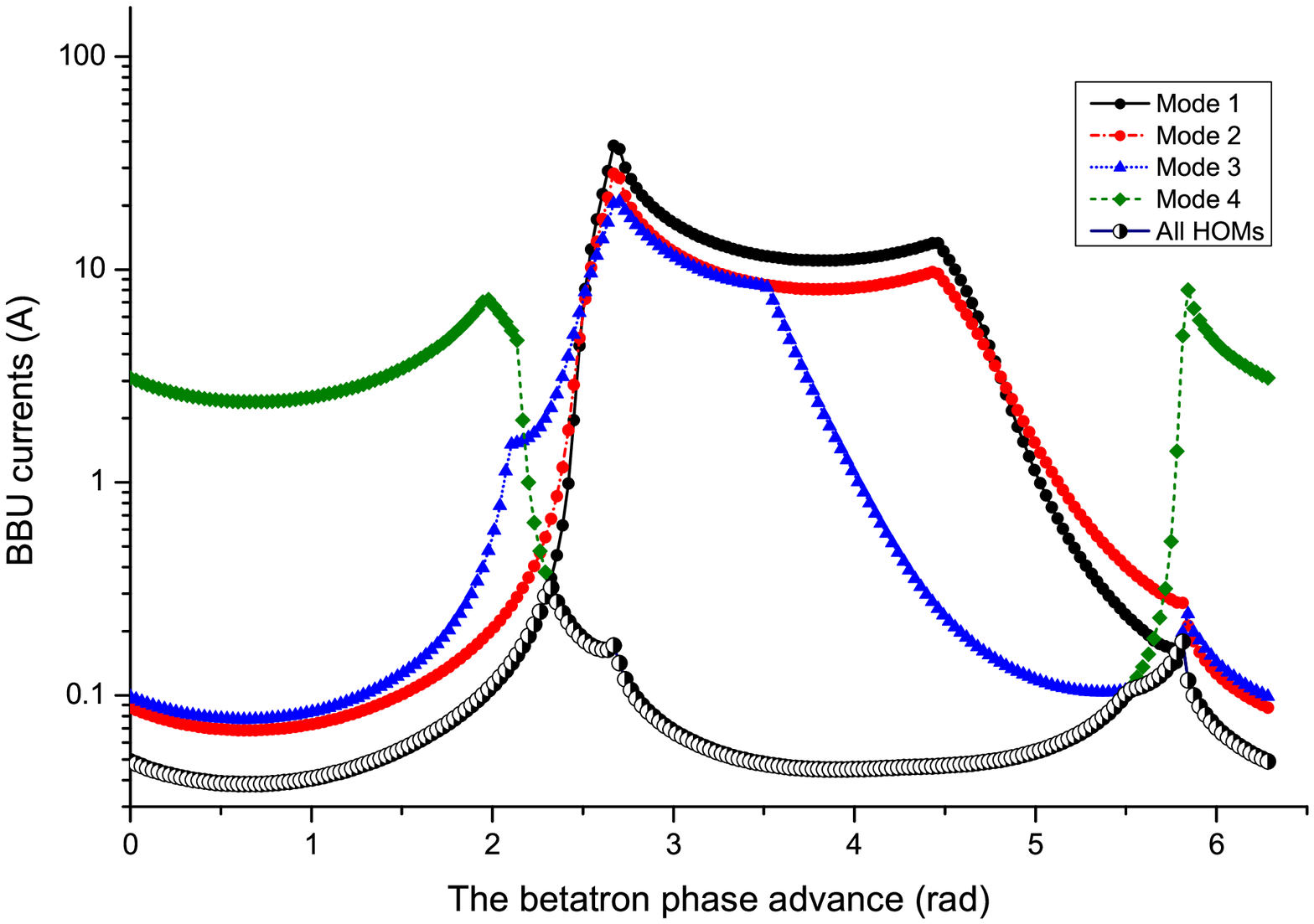}
\figcaption{\label{fig1}   The BBU current vs. the betatron phase advance of recirculating loop. 4 most threatening HOMs exist in each cavity.}
\end{center}

As shown in Fig.~(\ref{fig1}), the most threatening modes in the 9-cell Tesla cavity are $Mode$ 1 and $Mode$ 4. Both of them have larger $(R/Q)Q_e{/f}$ than other HOMs and they determine the threshold current of 9-cell Tesla cavity. The BBU current due to some HOMs is sensitive to the betatron phase advance so that a slight shift of betatron phase advance can lead to an obvious change of BBU current. The maximum value of BBU current that can be achieved by lattice adjustment for this case is about 300 mA and the minimum value is about 35 mA so that the threshold current for this case should be about 300 mA.

For an ERL with higher energy, more 9-cell Tesla cavities are required. Along with the increasing number of cavities, the electron beam will suffer more kicks and the offset after recirculating will be larger so that more energy exchange will occur between HOMs and the beam. We set ERLs with different number of cavities and calculate their BBU current. The simulation results are shown in Fig.~(\ref{fig2}).

\begin{center}
\includegraphics[width=8.0cm]{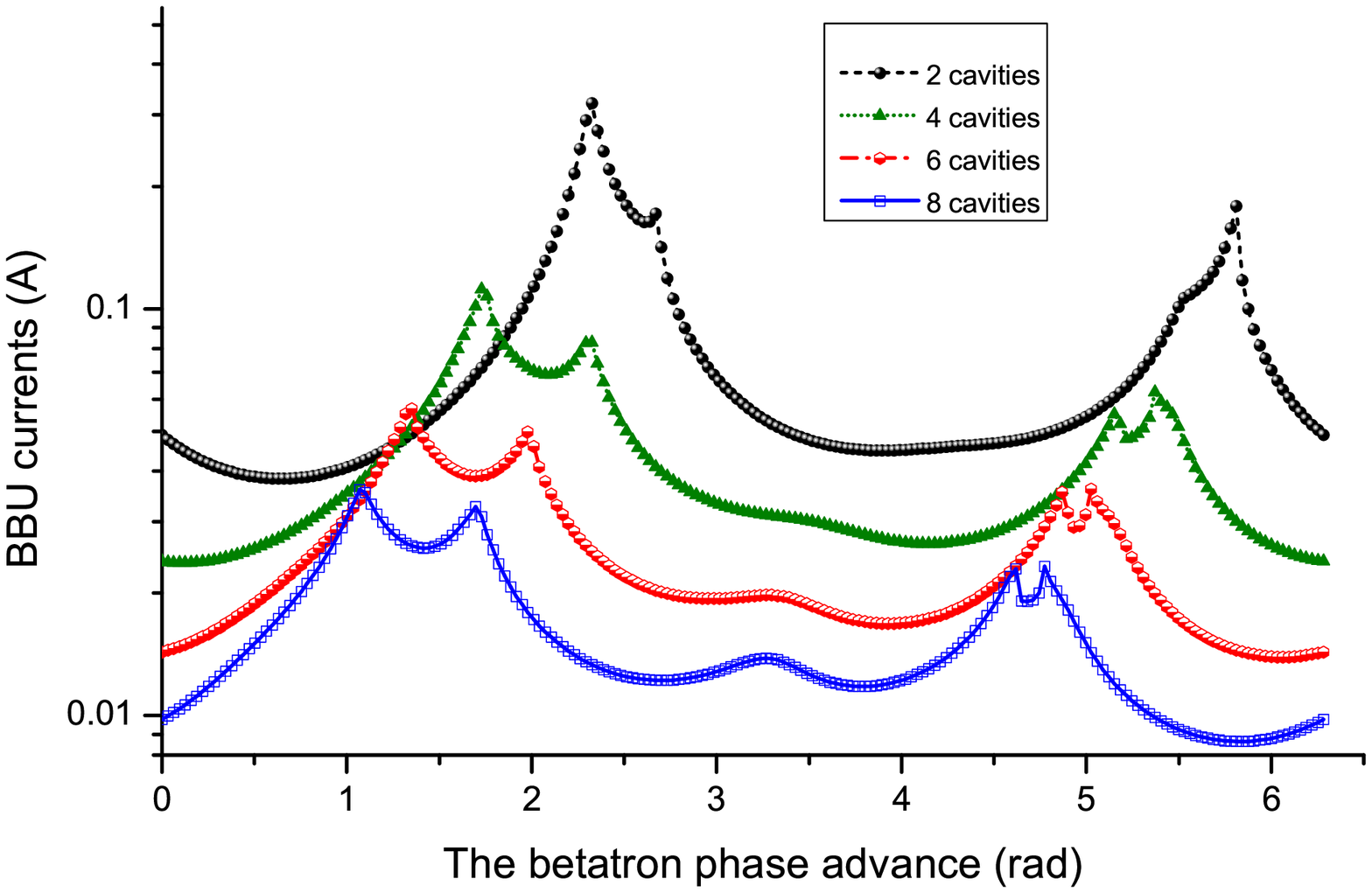}
\figcaption{\label{fig2}   The BBU current vs. the betatron phase advance of recirculating loop for different cavity numbers. The injection energy is 4 MeV.}
\end{center}

From Fig.~(\ref{fig2}) we can find that for the case of 8 cavities (in an ILC cryomodule) accelerating injected beam from 4 MeV to 100 MeV (square line in Fig.~(\ref{fig2})), the BBU threshold current is just about 31 mA. That means if we want to apply 9-cell Tesla cavity to such a scheme with average current higher than 31 mA, some additional methods should be considered.

\section{Influence of inhomogeneous HOMs to BBU}

During the fabrication of Tesla cavities, some errors and uncertainties are inevitable. These errors will make the dipole HOMs in real cavities slightly differ from the same HOM in ideal cavities. According to the former study \cite{lab10}, the frequency spread of the dipole HOMs due to the fabrication error is of the order of 10 MHz comparing with the ideal cavity. For an ERL with several Tesla cavities, a frequency spread of the same HOMs between different cavities will be introduced. This frequency spread may interrupt the coupling of HOMs in different cavities and increase the BBU current. Fig.~(\ref{fig3}) and Fig.~(\ref{fig4}) shows the BBU current vs. HOM frequency spread of mode1 in an 8-cavity scheme. The optics was chosen correspondingly to both the minimum and maximum current values from Fig.~(\ref{fig2}) and frequency has uniform distribution..

\begin{center}
\includegraphics[width=7.5cm]{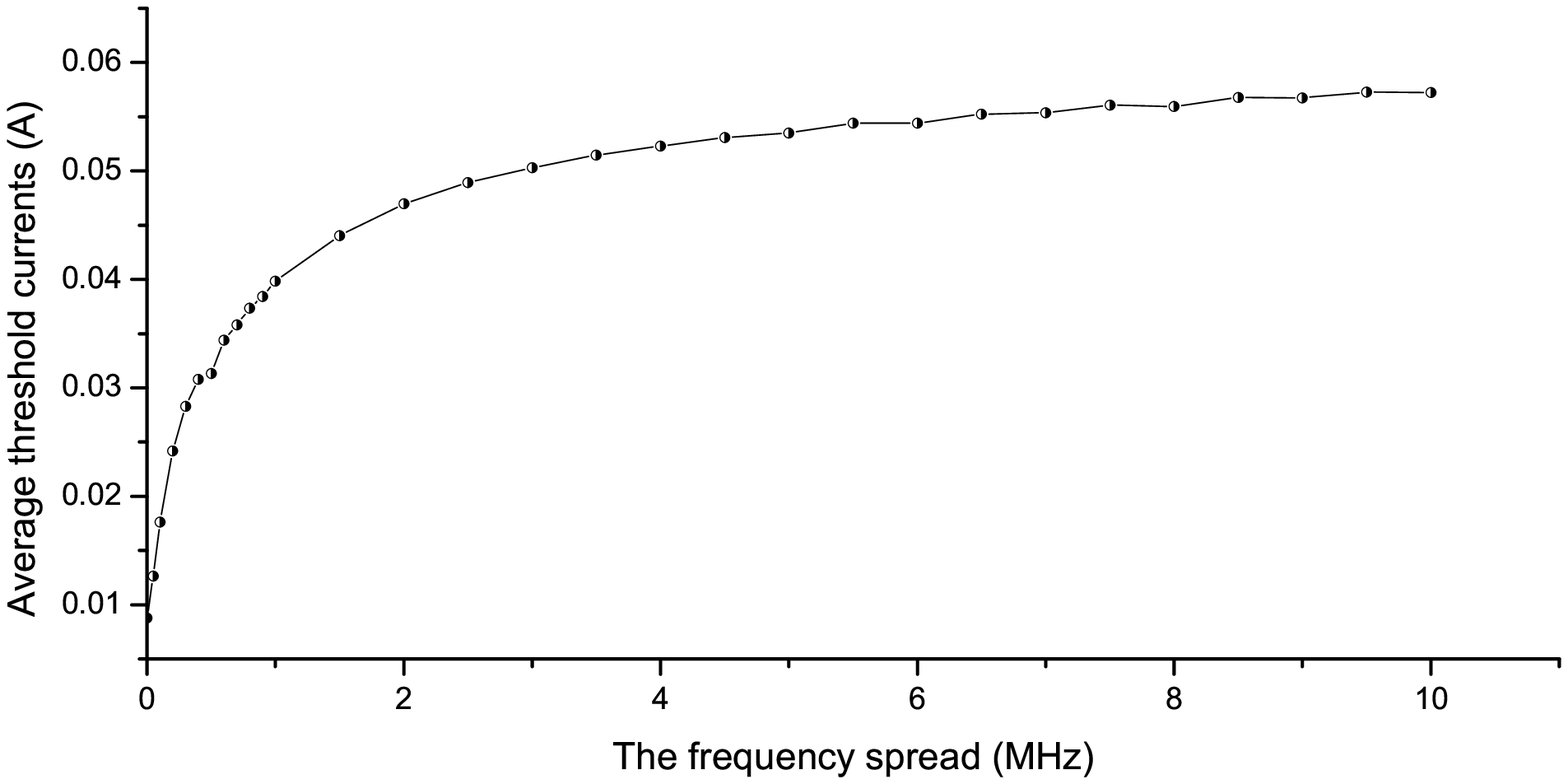}
\figcaption{\label{fig3}   The BBU current vs. the frequency spread. The lattice corresponds to the minimum value of BBU current $I_{th}=8.7$ mA}
\includegraphics[width=7.5cm]{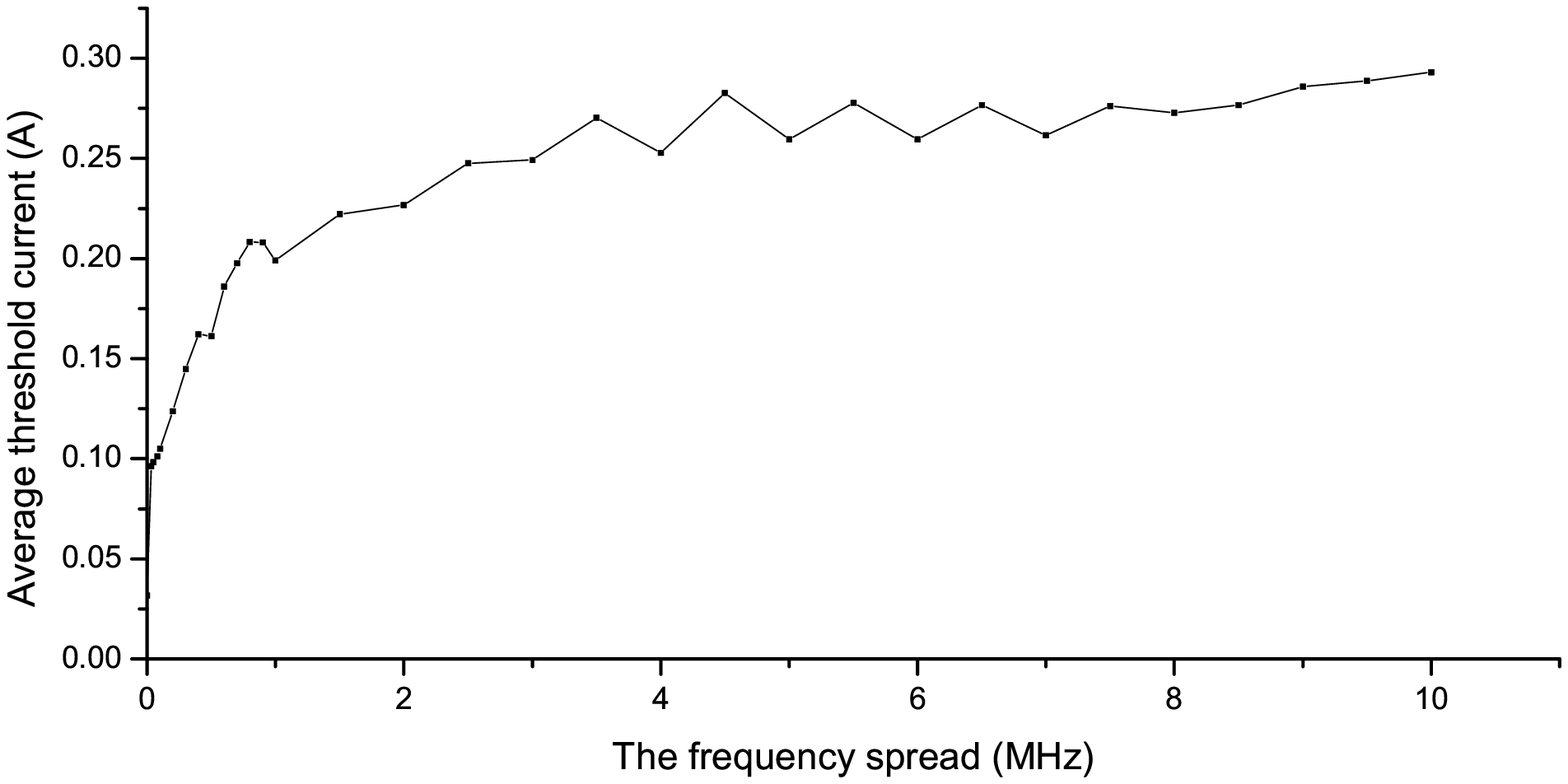}
\figcaption{\label{fig4}   The BBU current vs. the frequency spread. The lattice corresponds to the maximum value of BBU current $I_{th}=31 $ mA}
\end{center}

Clearly the worst scenario of BBU is the case that all cavities have the same HOM frequency. The HOM frequency spread between various cavities leads to a several times larger BBU current for $\sigma>3.5$ MHz, reaching about 50 mA for this case. At the same time we can find that when $\sigma>3.5$ MHz, the BBU current does not increase so fast as $\sigma<3.5$ MHz. That means the ability of increasing BBU current by HOM frequency spread is limited.

The distribution of HOM frequency is random so that its effect on BBU current is also random. Fig.~(\ref{fig5}) shows the statistics of BBU current against different cases of frequency spread in the 8-cavity scheme. The HOM frequency spread behaves a uniform distribution with $\sigma=10$ MHz around $Mode$ 1. The BBU current is calculated 5000 times and the average current of this case is about 60 mA, corresponding to the original value of BBU current $I_{th}=8.7$ mA.

\begin{center}
\includegraphics[width=7.5cm]{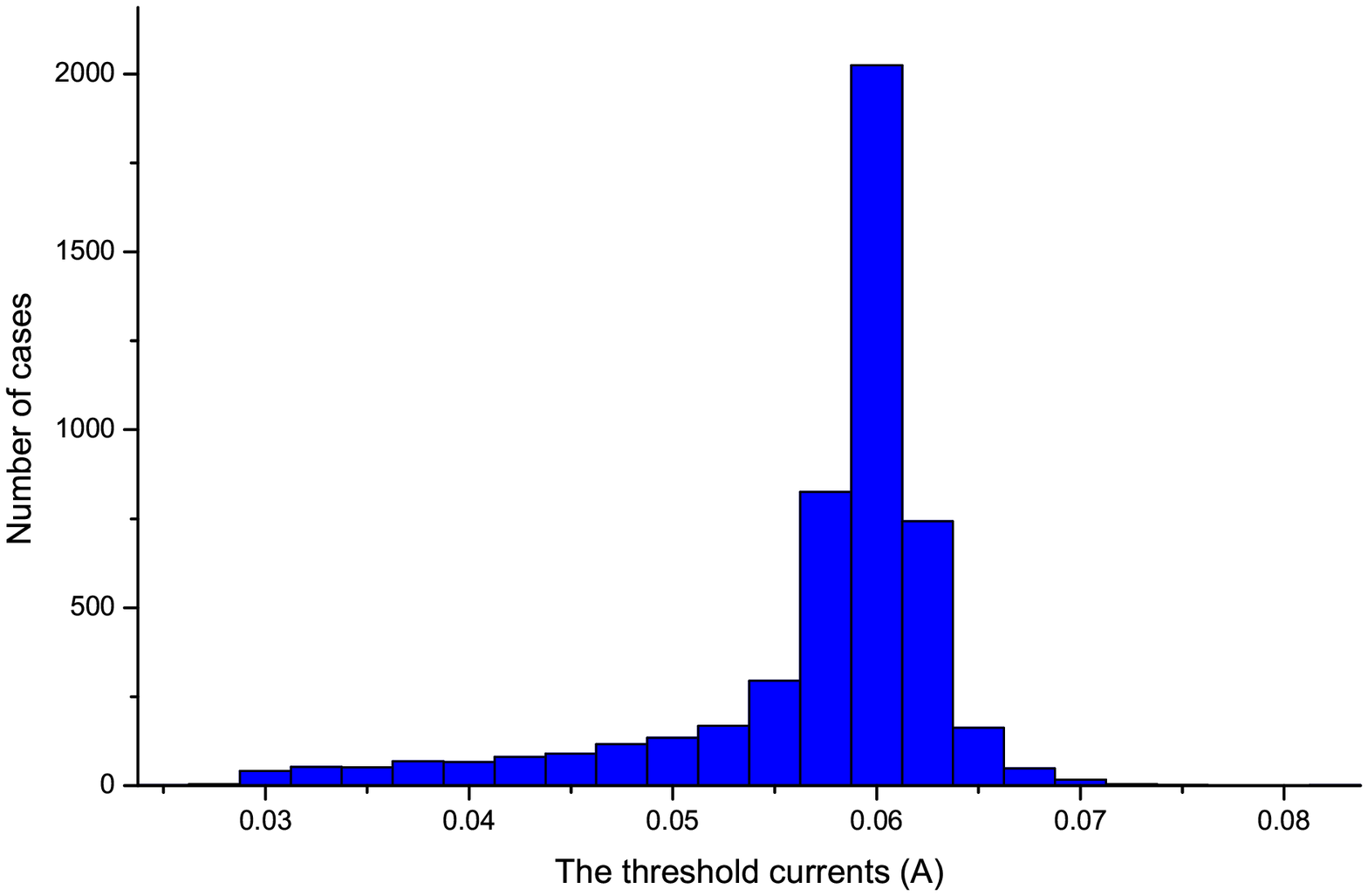}
\figcaption{\label{fig5}   Statistics of the BBU current for 10 MHz frequency spread.}
\end{center}

Except the frequency spread, cavity fabrication error will also introduce $Q_e$ spread of HOMs. e.g., slight adjustment of the depth of HOM coupler antenna insert into the cavity will cause an obvious change on $Q_e$. The shift of $Q_e$ due to cavity assembling uncertainties might be as large as one order of magnitude. Fig.~(\ref{fig6}) shows the statistics of the BBU current against the $Q_e$ spread. The BBU currents were calculated by determining the thresholds in 1000 random seeds that have $Q_e$ randomly distributed between $2.5\times10^4$ and $2.5\times10^5$. The betatron phase advance of lattice was chosen, which corresponds to the lowest value of BBU current 8.7mA. The average BBU current of the statistical result is about 3.1mA.

\begin{center}
\includegraphics[width=7.5cm]{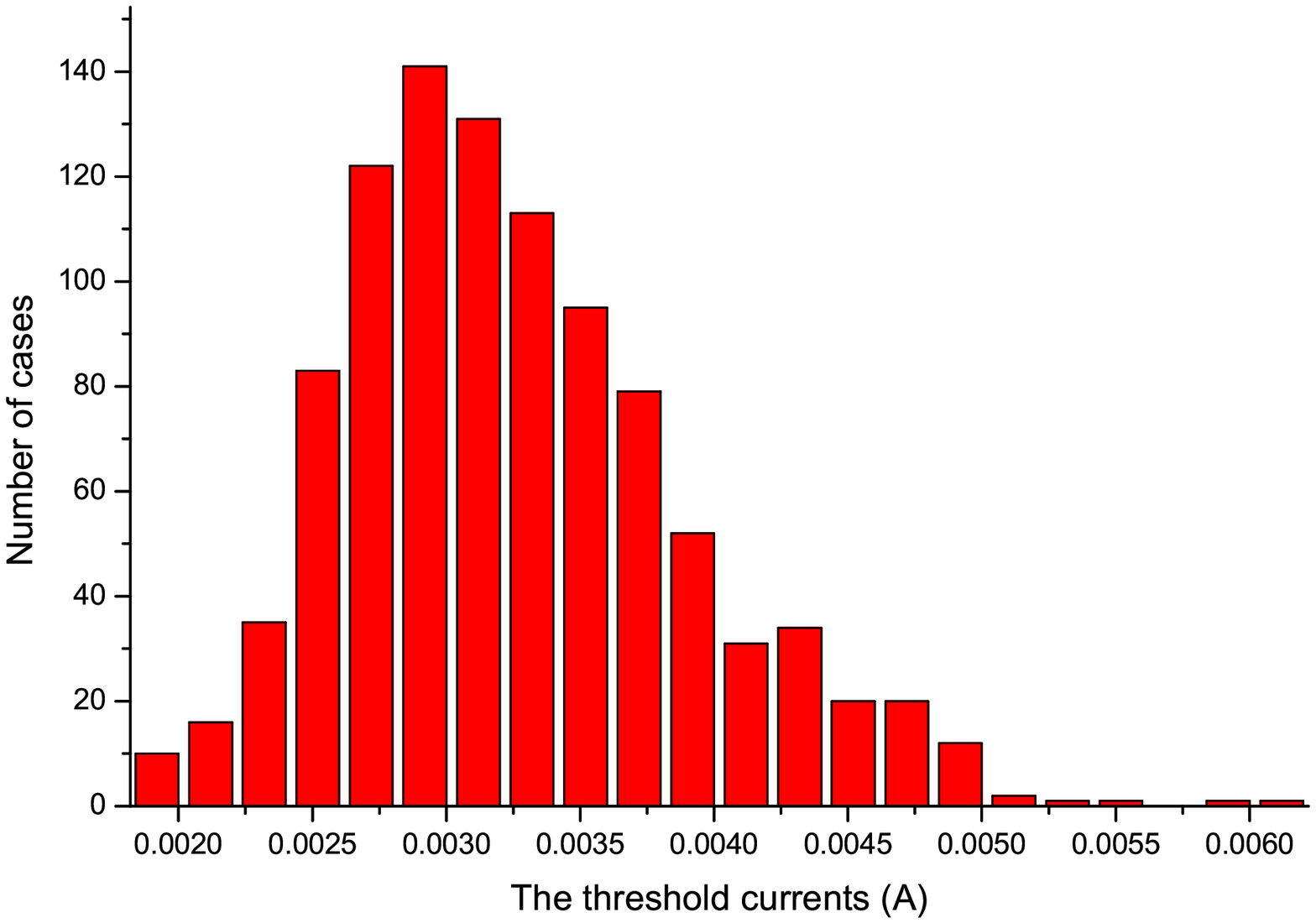}
\figcaption{\label{fig6}   Statistics of the BBU current for different cases of $Q_e$ distribution. HOM parameters: $f\approx$1.7074GHz, $R/Q=87.54\Omega$, $Q_{e}\in(2.5\times10^4,2.5\times10^5)$}
\end{center}

Typically the fabrication uncertainties of cavities will cause the $Q_e$ of HOM larger than nominal value so that the BBU current will be smaller than that of the ideal cavities. From Eq.~(\ref{eq1}), it can be seen that cavities at lower energy section contribute more to the instability because bunches are easier to be deflected in those cavities. Therefore, it is beneficial to set the first and last cavities in ERL to have well damped $Q_e$.

\section{Methods to suppress BBU}

According to Eq.~(\ref{eq1}), a smaller $(R/Q)Q_e$ indicates a higher threshold current so that the best approach of suppressing beam breakup is to fabricate the cavities with sufficient HOM damping. For ERLs with 9-cell Tesla cavities, as shown above, several not-well-damped HOMs may increase the risk of beam breakup instability of ERLs. Some methods can be applied to reduce this risk, such as a random frequency distribution introduced to HOMs among cavities \cite{lab11}, or a dedicated section to adjust the betatron phase advance of recirculating loop \cite{lab12}. For the former method, as shown in Fig.~(\ref{fig3}) and Fig.~(\ref{fig4}), the ability of increasing the threshold current is limited. For the latter method, we cannot make sure the value of $M_{12}^{*}\sin{\omega{T_r}}\approx0$ for each HOM in each cavity so that for ERLs with more cavities and HOMs the ability of suppressing BBU is also limited.

Except that, another two methods for beam optics control of BBU can also be applied, one is a reflection transport matrix which interchanges the horizontal and vertical planes betatron motion, while the other one is a rotation matrix which rotates the betatron phase plane by $90^{\circ}$ \cite{lab4}. These functions can be realized by a solenoid or a set of skew-quadruples inserted in the beamline. The coupled transport matrixes make the $T_{12}$ and $T_{34}$ of transport matrix zero so that $M_{12}^{*}=0$. The reflection matrix $M_{ref}$ and rotation matrix $M_{rot}$ can be expressed as

\begin{eqnarray}
\label{eq6}
M_{ref}=
\begin{pmatrix}
  0 & I\\
  I & 0\\
\end{pmatrix},
M_{rot}=
\begin{pmatrix}
  0 & I\\
  -I & 0\\
\end{pmatrix}
\end{eqnarray}

$I$ is the 2$\times$2 identity matrix. We insert a reflection section and a rotation section into the recirculating beamline of a 8-cavity scheme respectively and calculate their BBU current. The simulation results of these two configurations are shown in Fig.~(\ref{fig7}).

\begin{center}
\includegraphics[width=7.5cm]{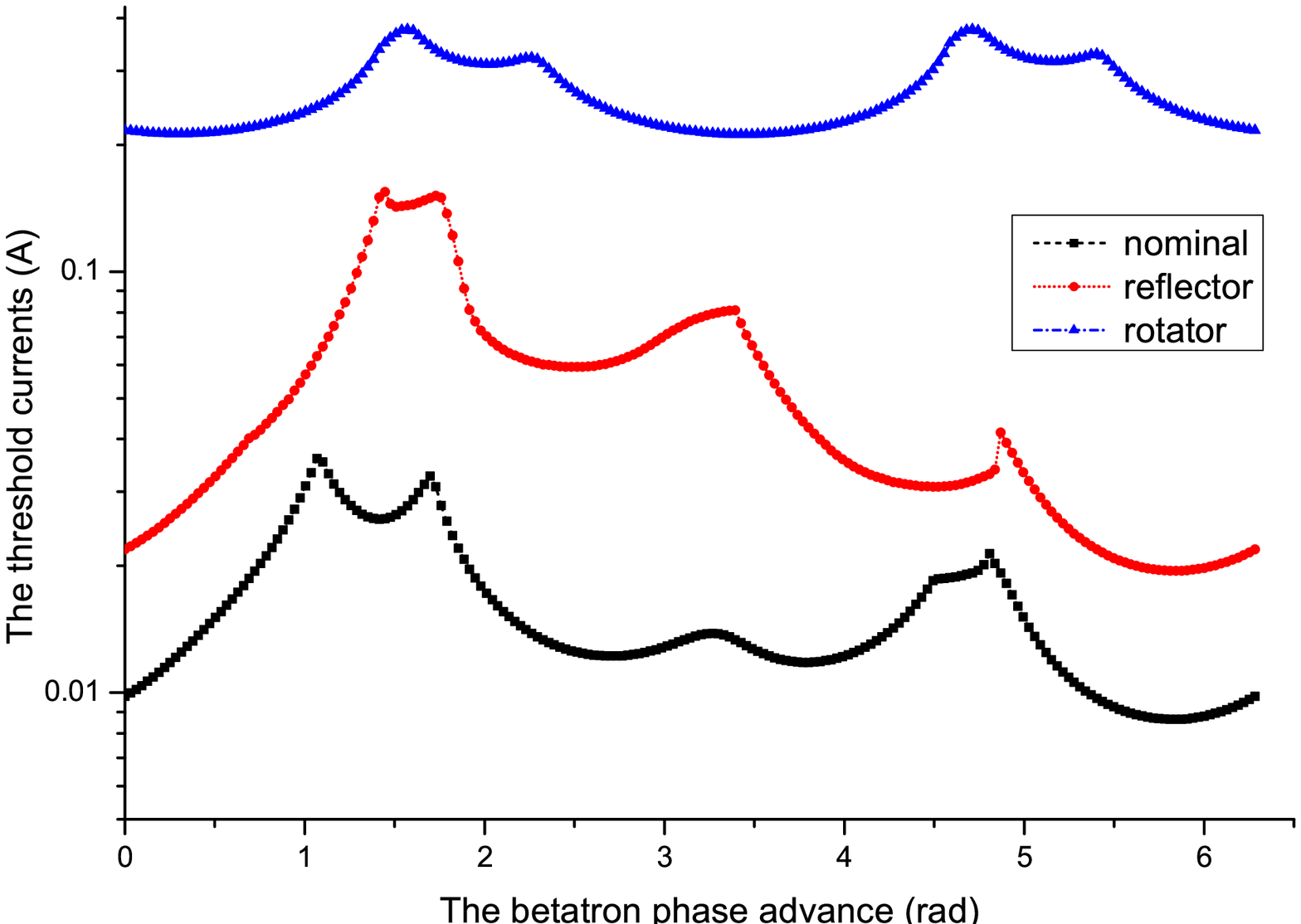}
\figcaption{\label{fig7}   The 8$\times$9cell Tesla cavities scheme with a reflector (red squares) or a rotator (green triangle) in the transport line.}
\end{center}

As shown in Fig.~(\ref{fig7}), a reflector or rotator in transport line would increase the threshold current obviously. For the reflection matrix, the BBU threshold is increased by a factor of about 5 and for the rotation matrix the factor is about 10. Theoretically these methods will lead to an infinite threshold current for single HOM in single cavity. However, for larger ERLs with more cavities and cryomodules, more complicated situation of HOMs may lead to more destructive mode coupling and degrade the performance of suppression. What's more, for ERLs of more than 2 turns, the coupling induced by these two methods will increase the difficulty of beam transportation.

\section{Conclusion}

The BBU threshold currents of compact ERLs with 9-cell Tesla cavities are investigated. The study shows that by adjusting the betatron phase advance of recirculating lattice and introducing frequency spread between different cavities, the BBU threshold current up to hundreds mA can be obtained for an ERL test facility with two 9-cell Tesla cavities, which is sufficient for the requirement of PKU-ERL test facility. For an ERL test facility with $8\times9$-cell Tesla cavities, the BBU threshold current up to tens mA can be obtained, too. It is feasible to use 9-cell Tesla cavity on some compact ERL test facilities with just a few cavities and beam current around tens mA.

\section{Acknowledgment}

This work was supported by the Major State Basic Research Development Program of China under Grant No. 2011CB808303 and No. 2011CB808304.

\end{multicols}

\vspace{-1mm}
\centerline{\rule{75mm}{0.1pt}}
\vspace{2mm}

\begin{multicols}{2}

\end{multicols}

\clearpage

%\end{CJK*}
\end{document}